\documentstyle[preprint,tighten,version2,aps]{revtex}

{\makeatletter%
\gdef\labeleqs#1{{%
\edef\@currentlabel{%
\ifappendixon\appletter\fi
\ifsecnumbers\ifnum\c@secnum>0
\arabic{secnum}.\fi\fi\arabic{equation}}%
\label{#1}%
}}%
}
\begin{document}
\draft
\preprint{IFUP-TH 3/95}
\begin{title}
Topological charge density renormalization in the presence of dynamical
fermions.
\end{title}
\author{B. All\'es, A. Di Giacomo, H. Panagopoulos$^\star$, E. Vicari}
\begin{instit}
Dipartimento di Fisica dell'Universit\`a and I.N.F.N.,
I-56126 Pisa, Italy\\
$^\star$ Department of Natural Sciences, University of Cyprus,
Nicosia, Cyprus
\end{instit}
\begin{abstract}
We study the renormalization group behaviour of the
topological charge density in full QCD on the lattice. We propose a
way of extracting the necessary renormalization functions from Monte
Carlo simulations.
\end{abstract}


\narrowtext

One important issue which has come up in recent years, in the study of
strong interactions, is the so called proton spin
crisis~\cite{Carlitz,Ellis,Shore}. The issue concerns the on-shell
nucleon matrix element of the singlet axial current $j_\mu^5$
\begin{equation}
\langle\, \vec{p},e |\,j_\mu^5\,| \vec{p}\,', e'\,\rangle\;=\;
\bar{u}(\vec{p},e)\left[ G_1(k^2)\gamma_\mu \gamma_5\,-\,
G_2(k^2)k_\mu\gamma_5\right] u(\vec{p}\,',e')
\label{jmatrix}
\end{equation}
where $e,e'$ label the helicity states and $k$ is the momentum transfer.
In a naive wave function picture $G_1(0)$ can be interpreted as the
fraction of the nucleon spin carried by the quarks. Experimental
determinations lead to an unexpectedly small value of $G_1(0)$, calling for
a theoretical explanation in the context of a nonperturbative study in QCD.
$G_1(0)$ can be extracted from the nucleon matrix element
of the topological charge density $q(x)$
by using the fact that in the chiral limit
the anomalous divergence of $j_\mu^5$ is proportional to the
topological charge density. In the presence of dynamical fermions, one has
\begin{equation}
\langle\, \vec{p},e |\,q\,| \vec{p}\,', e'\,\rangle\;=\;
MB(k^2)\,\bar{u}(\vec{p},e)i\gamma_5 u(\vec{p}\,',e'),
\label{qp}
\end{equation}
with
\begin{equation}
N_f B(0) \;=\;G_1(0)\;.
\label{q2p}
\end{equation}
Preliminary lattice studies of the on-shell matrix element of $q(x)$, in
quenched QCD, have been done in Refs.~\cite{Gupta,Mandula,spin}.
An unquenched study using a geometrical definition of topological charge
density
has been reported in Ref.~\cite{Schierholz}. We also would like to note
the direct calculations of the on-shell proton $j_\mu^5$ matrix element
performed in quenched QCD reported in Refs.~\cite{Fukugita,Dong}.

Notice that the above discussion overlooks the non-invariance
under the renormalization group of the operators involved;
indeed, unlike pure gauge theory, in full QCD the topological charge density
mixes under renormalization with $\partial_\mu j_\mu^5$.
Therefore the relations of the matrix elements of $j_\mu^5$ and $q(x)$
with the experimental
observables like $G_1(0)$ cannot be strictly given by
Eqs.~(\ref{jmatrix}) and (\ref{q2p}). Instead, one must look for
equations involving renormalization group invariant quantities. This
will appear
clearer in the following, when we analyze the renormalization of operators
regularized on the lattice.

In this letter we address the scale dependence of the matrix
elements of operators entering the anomaly equation, in lattice
regularization, and provide the prescriptions necessary to undertake a
lattice calculation of $G_1(0)$ in full QCD.

Let us first discuss this issue in the case of dimensional regularization with
minimal subtraction. In euclidean space
\begin{equation}
\partial_\mu j_\mu^5(x)\;=\; i2N_f q(x) + i 2p(x)\;,
\label{anomaly}
\end{equation}
where
\begin{eqnarray}
j_\mu^5(x)&=& \sum_{f=1}^{N_f}
\bar{\psi}_f(x)\gamma_\mu\gamma_5\psi_f(x)\;,\nonumber \\
q(x)&=& {g^2\over 64\pi^2} F_{\mu\nu}^a (x)\tilde{F}_{\mu\nu}^a(x)\;,\nonumber
\\
p(x)&=& \sum_{f=1}^{N_f} m_f\bar{\psi}_f(x)\gamma_5\psi_f(x)\;.
\label{oper}
\end{eqnarray}
The nonrenormalizability property of the anomaly in the ${\overline {\rm MS}}$
scheme means that the anomaly equation should take exactly the same form in
terms
of bare or renormalized quantities. However the renormalization of the three
pseudoscalar gauge invariant operators $\partial_\mu j^5_\mu(x)$, $q(x)$ and
$p(x)$ turns out to be nontrivial; in the ${\overline {\rm MS}}$
scheme it reads~\cite{Espriu,Veneziano}
\begin{equation}
\left( \matrix{  i2N_f\,q(x)\cr \partial_\mu j_\mu^5(x)\cr i2p(x)\cr}\right)^R
\;=\
\left( \matrix{ 1&z-1&0\cr 0&z&0\cr 0&0&1\cr}\right)\;
\left( \matrix{  i2N_f\,q(x)\cr \partial_\mu j_\mu^5(x)\cr i2p(x)\cr}\right)^B
\label{cren}
\end{equation}
where, to order $g^4$,
\begin{equation}
z\;=\; 1 \,-\, {g^4\over 16\pi^4} {3c_{_F}\over 8} \, N_f \, {1\over
\epsilon}\,+\,
O\left( g^6\right)\;,
\label{zeta}
\end{equation}
$\epsilon = d/2 -2$ and $c_{_F}={N^2-1\over 2N}={4\over 3}$.
The structure of the renormalization matrix above assures the stability of the
anomaly equation
under renormalization. The associated anomalous dimension is given by
\begin{equation}
\gamma_{_{\overline {\rm MS}}}(g)\;=\;\mu {{\rm d} Z\over {\rm d}\mu}
Z^{-1}\;=\;
\left( \matrix{ 0&\bar{\gamma}(g)&0\cr 0&\bar{\gamma}(g)&0\cr
0&0&0\cr}\right)\;,
\label{candim}
\end{equation}
where
\begin{equation}
\bar{\gamma}(g)\;=\;
\mu {{\rm d} \over {\rm d}\mu} \ln z \;=\;
\gamma_2 \,g^4\,+\, \gamma_3 \,g^6 \,+\,O\left( g^8\right)\;,
\label{candim2}
\end{equation}
and
\begin{equation}
\gamma_2\;=\;- {1\over 16\pi^4} {3c_{_F}\over 2} \, N_f \;.
\end{equation}

We consider now the lattice regularization of QCD given by the Wilson action.
Bare operators are defined as local analytic functions of elementary
fields, i.e. $U_\mu(x)$ and $\psi(x)$. At finite lattice spacing
the regularized version of a continuum operator is not unique, indeed
regularized versions of the same continuum operator  can differ
by $O(a)$ terms (at the classical level). We consider the following lattice
regularizations of
$j_\mu^5(x)$, $\partial_\mu j_\mu^5(x)$ and $q(x)$:
\begin{equation}
a^3 j^{5,L}_{\mu}\;=\; {1\over 2} \left[
\bar{\psi}(x+\mu)\gamma_\mu \gamma_5 U_\mu^\dagger(x) \psi(x) +
\bar{\psi}(x)\gamma_\mu \gamma_5 U_\mu (x) \psi(x+\mu)\right]\;,
\label{lattj5}
\end{equation}
\begin{equation}
a \Delta_\mu j^{5,L}_\mu\;=\; \left[ j^{5,L}_\mu (x) - j^{5,L}_\mu
(x-\mu)\right]\;,
\label{lattdj5}
\end{equation}
(these definitions come out naturally from the lattice Ward identities
associated to an axial vector flavour-singlet
transformation\cite{Karsten}), and
\begin{equation}
a^4 q^L\;=\;- {1\over 2^4\times 32 \pi^2}
\sum^{\pm 4}_{\mu\nu\rho\sigma=\pm 1}
\epsilon_{\mu\nu\rho\sigma} {\rm Tr}
\left[ \Pi_{\mu\nu}\Pi_{\rho\sigma}\right]
\label{Q^L}
\end{equation}
($\Pi_{\mu\nu}$ is the product of link variables around a
plaquette)~\cite{DiVecchia},
which has been largely used in quenched lattice QCD studies. The
renormalization of $q^L$ in pure gauge theory has been studied
perturbatively in
Ref.~\cite{Campo}; a nonperturbative study for SU(3) was performed in
Ref.~\cite{spin}.

The $r$-term in the Wilson action breaks explicitly
chiral invariance, which should be recovered at a non-trivial value of
the bare quark masses.
Perturbative calculations and general arguments support the existence of a
chiral limit, in which the axial anomaly should be
reproduced~\cite{Karsten,Bochicchio,Smit}.
In that limit, any mixings of $\Delta_\mu j^{5,L}_\mu$ and $q^L$ with lower
dimension operators,
in particular $\bar{\psi}\gamma_5 \psi$, must necessarily vanish, for
the anomaly equation to hold. Indeed, mixing with
$\bar{\psi}\gamma_5\psi$ should be completely absorbed in a
redefinition of renormalized quark masses; thus,
in the chiral limit (i.e. $m_\pi\rightarrow 0$), only the operators
$\Delta_\mu j^{5,L}_\mu$ and $q^L$ need be considered in a
renormalization analysis.
In the following we study the renormalization of
$\Delta_\mu j^{5,L}_\mu$ and $q^L$ in the chiral limit.
A more general analysis for non-zero renormalized mass values
(i.e. $m_\pi\neq 0$) requires the introduction of a regularized version
of $m\bar{\psi}\gamma_5\psi$, but this should not lead to more than few
straightforward
changes in the following considerations.

Since we are interested in the matrix elements of operators in the
${\overline {\rm MS}}$ scheme, we must renormalize our bare lattice
operators $j^{5,L}_\mu$ and $q^L$ in such a way that their
renormalized correlation functions coincide with those obtained in
${\overline  {\rm MS}}$.
This should be achieved by a lattice renormalization matrix $Z_L$:
\begin{equation}
\left( \matrix{ i2N_f\,q\cr \partial_\mu j_\mu^5\cr }\right)^R
\;=\
\left( \matrix{ Z_{qq}&Z_{q\psi}\cr Z_{\psi q}&Z_{\psi\psi}\cr}\right)\;
\left( \matrix{ i2N_f\,q^L\cr \Delta_\mu j_\mu^{5,L}\cr }\right)
\label{crenlatt}
\end{equation}
It is worth mentioning at this point that, unlike the case of
flavor-nonsinglet currents, a nonperturbative evaluation of the matrix
elements of $Z_L$ cannot be obtained by use of Ward identities
according to the method developed in Ref.~\cite{Bochicchio}.

Since the anomalous dimension matrix $\gamma_{_{\overline {\rm MS}}}(g)$
must depend only on the renormalization
scheme, and not on the type of regularization, we must have
\begin{equation}
\mu { {\rm d} Z_L \over {\rm d} \mu} Z^{-1}_L\;=\;
\gamma_{_{\overline {\rm MS}}}(g)\;.
\label{gflatt}
\end{equation}
This provides some constraints on the form of $Z_L$, in particular
on its logarithms. We should have that:
($i$) $Z_{qq}$ is a finite function of the bare lattice coupling $g_0$;
($ii$) The logarithms appear at order $g^4_0$ in $Z_{q\psi}$ and
$Z_{\psi\psi}$;
($iii$) $Z_{\psi q}=0$~\cite{Smit}.
Therefore
\begin{eqnarray}
Z_{qq}&=& 1+g_0^2 d_{qq}^{(1)}+g_0^4 d_{qq}^{(2)}+O\left(g^6_0\right)\nonumber
\\
Z_{q\psi}&=& g_0^4 \left( c_{q\psi}^{(2)} \ln a\mu + d_{q\psi}^{(2)}\right)
+O\left(g^6_0\right)\nonumber \\
Z_{\psi\psi}&=& 1+g_0^2 d_{\psi\psi}^{(1)}
+g_0^4 \left( c_{\psi\psi}^{(2)}\ln a\mu +
d_{\psi\psi}^{(2)}\right)+O\left(g^6_0\right)\;.
\label{zep}
\end{eqnarray}
Some of the above coefficients are known. It is easy to see that
$d_{qq}^{(1)}$ is equal to the corresponding coefficient of the multiplicative
renormalization of $q^L$ in a pure gauge theory~\cite{Campo,scaling2}
\begin{equation}
d_{qq}^{(1)}\;=\;N[-{1 \over {4 N^2}} + Z_0 + {1 \over 8} + {1 \over {2
\pi^2}}]\;,
\label{resold}
\end{equation}
where $Z_0=0.15493$. $d_{qq}^{(1)}\simeq 0.9084$ for $N=3$.
{}From Eq.~(\ref{gflatt})
\begin{equation}
c\;\equiv\;c_{q\psi}^{(2)}\;=\;c_{\psi\psi}^{(2)}\;=\;-{1\over
  16\pi^4} {3c_{_F}\over 8}\, N_f\;.
\label{cpsipsi}
\end{equation}
$d_{\psi\psi}^{(1)}$ depends on the Wilson parameter $r$; for $r=1$,
we find, using the results of Ref.~\cite{Martinelli},
$d_{\psi\psi}^{(1)}\simeq -0.073\,$.

In the following we set
\begin{equation}
O_R(\mu) = \left( \matrix{ i2N_f\,q\cr \partial_\mu j_\mu^5\cr }\right)^R\;,
\;\;\;\;\;\;\;\;\;\;\;\;\;
O_L = \left( \matrix{ i2N_f\,q^L\cr \Delta_\mu j_\mu^{5,L}\cr }\right)\;.
\label{def}
\end{equation}
Under a change of scale, the matrix elements of $O_R(\mu)$ satisfy
\begin{equation}
\mu{{\rm d} \over {\rm d} \mu} \langle O_R\rangle \;=\;
\gamma_{_{\overline {\rm MS}}} \; \langle O_R\rangle\;,
\label{requ}
\end{equation}
whose formal solution is
\begin{equation}
\langle O_R \rangle (\mu^\prime)\;=\;\left[ {\cal P} \exp
\int^{g(\mu^\prime)}_{g(\mu)}
{\gamma_{_{\overline {\rm MS}}} (\tilde{g})
\over \beta_{_{\overline {\rm MS}}}(\tilde{g})} {\rm d} \tilde{g}\right]
\langle O_R \rangle (\mu)\;,
\label{rsol}
\end{equation}
where $\beta_{_{\overline {\rm MS}}}(g)$ is the $\beta$-function in
the ${\overline {\rm MS}}$ scheme and the symbol $\cal P$ indicates
path ordering. Given the particular form of the matrix
$\gamma_{_{\overline {\rm MS}}}$ in our case, it is easy to see that
$\cal P$ may be dropped.
Proceeding as in Ref.~\cite{scaling1}, we choose $\mu^\prime=1/a$ and write
\begin{equation}
\langle O_R \rangle (1/a)\;=\;Z_L(1/a) \langle O_L \rangle \;=\;
\left[ \exp \int^{g(1/a)}_0
{\gamma_{_{\overline {\rm MS}}} (\tilde{g})
\over \beta_{_{\overline {\rm MS}}}(\tilde{g})} {\rm d} \tilde{g}\right]
\left[ \exp \int^0_{g(\mu)}
{\gamma_{_{\overline {\rm MS}}} (\tilde{g})
\over \beta_{_{\overline {\rm MS}}}(\tilde{g})} {\rm d} \tilde{g}\right]
\langle O_R \rangle (\mu)\;,
\end{equation}
and therefore
\begin{equation}
\langle O_L \rangle \;=\;\left[ Z_L(1/a)^{-1}\;
 \exp \int^{g(1/a)}_0
{\gamma_{_{\overline {\rm MS}}} (\tilde{g})\over \beta_{_{\overline {\rm
MS}}}(\tilde{g})}
{\rm d} \tilde{g}\right]
\left[ \exp \int^0_{g(\mu)}
{\gamma_{_{\overline {\rm MS}}} (\tilde{g})\over
\beta_{_{\overline {\rm MS}}}(\tilde{g})} {\rm d} \tilde{g}\;
\langle O_R \rangle (\mu)\right]\;.
\end{equation}
$g(\mu)=Z^L_g(\mu) g_0$, where $g(\mu)$ is the coupling renormalized in
${\overline {\rm MS}}$~\cite{Hasenfratz,Kawai}.

Notice that the vector
\begin{equation}
M(\mu)\langle O_R \rangle (\mu)\;\equiv\; \exp \int^0_{g(\mu)}
{\gamma_{_{\overline {\rm MS}}} (\tilde{g})\over
\beta_{_{\overline {\rm MS}}}(\tilde{g})} {\rm d} \tilde{g}\;
\langle O_R \rangle (\mu)
\label{cdep}
\end{equation}
is renormalization group invariant, that is independent of $\mu$,
and it is what can be naturally extracted also from physical experiments.
The lattice dependence is restricted to
\begin{equation}
L(g_0)\;\equiv\; Z_L(1/a)^{-1}\;
 \exp \int^{g(1/a)}_0
{\gamma_{_{\overline {\rm MS}}} (\tilde{g})
\over \beta_{_{\overline {\rm MS}}}(\tilde{g})} {\rm d} \tilde{g}\;.
\label{ldep}
\end{equation}

The matrix $M(\mu)$
can be expanded in powers of $g(\mu)$:
\begin{equation}
M(\mu)\;=\;
\left( \matrix{ 1& {c\over 2b_0}g(\mu)^2\cr 0& 1+{c\over 2b_0}g(\mu)^2
\cr}\right)\;+
\;O\left( g(\mu)^4 \right)\;,
\label{mmu}
\end{equation}
where $b_0$ is the first coefficient of the $\beta$-function:
\begin{equation}
b_0\;=\;{1\over 16\pi^2}\left( {11\over3}N-{2\over 3}N_f\right)\;.
\end{equation}
Notice that
\begin{equation}
{c\over 2b_0}\;=\;-{N_f\over\pi^2}\left( {11\over 3}N - {2\over
3}N_f\right)^{-1}
\label{cmu}
\end{equation}
is a small number: $\simeq -0.0486$ for $N=3$ and $N_f=4$.

$L(g_0)$ can be expanded in powers of $g_0$; to order $g_0^2\,$:
\begin{equation}
L(g_0)\;=\;
\left( \matrix{ 1-d_{qq}^{(1)}g_0^2 & -{c\over 2b_0}g_0^2\cr
0& 1-\left(d_{\psi\psi}^{(1)}+{c\over 2b_0}\right)g_0^2 \cr}\right)\;+
\;O\left( g_0^4 \right)\;.
\label{lg0}
\end{equation}

Use of the anomaly equation for the renormalized operators simplifies
considerably the above formulae; indeed, in the chiral limit,
\begin{equation}
\partial_\mu j_\mu^5(x)^R \;=\; i2N_f q(x) ^R\;
\label{anren}
\end{equation}
implying
\begin{equation}
M(\mu)\langle O_R \rangle (\mu)\;=\;
\left( \matrix{
\langle T \rangle \cr
\langle T \rangle \cr} \right)
\label{Msim}
\end{equation}
where
\begin{equation}
\langle T \rangle \;\equiv\;
\langle\; \partial_\mu j_\mu^5(x)^R\;\rangle
\exp \int^0_{g(\mu)}
{\bar{\gamma}(\tilde{g})\over \beta_{_{\overline {\rm MS}}}(\tilde{g})} {\rm d}
\tilde{g}
\label{tt}
\end{equation}
We recall that $\bar{\gamma}(g)$ has been defined in Eq.~(\ref{candim}).
Thus, measuring  the matrix element $\langle q^L\rangle$
by Monte Carlo simulation, we obtain
\begin{equation}
\langle i2N_f q^L\rangle\;=\;
z_q(g_0^2)\,\langle T \rangle \;,
\label{ql}
\end{equation}
where
\begin{equation}
z_q (g_0^2)\;=\; L_{11}+L_{12}\;.
\end{equation}
Notice that $\langle T \rangle$ is renormalization group invariant.

{}From Eq.~(\ref{lg0}) we find $z_q(g_0^2)=1-0.8598 g_0^2 + O(g_0^4)$ for $N=3$
and $N_f=4$.
We recall that the typical values of $g_0^2$ where simulations on today's
supercomputer
can be performed are actually not small:
$g_0^2\simeq 1$.  Therefore as a consequence of the large $O(g_0^2)$
coefficient,
more  terms turn out to be  necessary to obtain an estimate
of $z_q(g_0^2\simeq 1)$ from perturbation theory.
Of course an improved lattice topological charge density operator having a
smaller $O(g_0^2)$
term in $z_q(g_0^2)$ would be welcome.
The evaluation of the $O(g_0^4)$ term in $z_q(g_0^2)$ needs the calculation of
the coefficients
$d_{qq}^{(2)}$ and $d_{q\psi}^{(2)}$ in Eqs.~(\ref{zep}) and $\gamma_3$ in
Eq.~(\ref{candim2}).
$d_{qq}^{(2)}$ requires a two loop lattice calculation. However an estimate may
be obtained
from the nonperturbative determination of the multiplicative renormalization
of $q^L$ in quenched QCD, performed in Ref.~\cite{spin}, supplemented with
relatively easier
analytical two loop calculations  of the only graphs containing fermion lines
among those contributing to the two-point gluonic function with an insertion of
$q^L$.
The evaluation of the other coefficients should be simpler,
$d_{q\psi}$ can be obtained from a one loop calculation, while
$\gamma_3$ requires a two loop calculation in dimensional regularization.

Given the difficulties of a perturbative evaluation, a nonperturbative
estimate of $z_q(g_0^2)$ would be welcome. To this purpose, notice that
the result (\ref{ql}) is  independent of the particular
matrix element. We may therefore eliminate the lattice factor
$z_q(g_0^2)$, masking the physical signal, by measuring two different matrix
elements of $q_L$.
For example, beside the on-shell nucleon matrix element, we may also
extract the matrix element $\langle \,0|\,q^L\,|\eta^\prime\,\rangle$
from a lattice calculation.
The wall-wall correlation function $G^{\rm w}_q(t)$ constructed from
$\langle \,q^L(x) q^L(y)\,\rangle$ should have the following
long distance behavior
\begin{equation}
G^{\rm w}_q(t)\;\simeq\;
{|\langle \,0|\,q^L\,|\eta^\prime\,\rangle |^2\over 2 m_{\eta^\prime}}
e^{-m_{\eta^\prime} t}
\label{gw}
\end{equation}
Furthermore
\begin{equation}
\langle \,0|\,i2 N_f  q^L\,|\eta^\prime\,\rangle\;=\;
z_q\,\langle \,0|\,T\,|\eta^\prime\,\rangle\;,
\label{ql2}
\end{equation}
and
\begin{equation}
\langle \,0|\,T\,|\eta^\prime\,\rangle\;=\;\sqrt{N_f} m^2_{\eta^\prime}
f_{\eta^\prime}\;.
\label{feta}
\end{equation}
In the large-$N$ limit
$f_{\eta^\prime}=f_{\pi}$~\cite{Witten}.
A more precise phenomenological estimate gives
$f_{\eta^\prime} \simeq 0.8 f_\pi$~\cite{etaprime,espe}.

We have thus seen how a Monte Carlo measurement of the proton matrix
element of $q^L$, in the presence of dynamical fermions, can be
related to the physical, {\it renormalization group invariant}
quantity relevant for the spin crisis problem; the renormalization
function involved in this process can be determined nonperturbatively
from measurements of $\langle \,q^L(x) q^L(y)\,\rangle$ and an
estimate of $f_{\eta^\prime}$.

\bigskip\noindent
{\bf Acknowledgements} We would like to thank G. Paffuti for helpful
conversations.

\end{document}